\documentclass[aps,showpacs,preprintnumbers,amsmath,amssymb,nofootinbib,epsf]{revtex4}
\usepackage{graphicx}

\begin{document}

\preprint{gr-qc/0503072}

\title{Conformally symmetric vacuum solutions of the gravitational field equations in the brane-world models}

\author{T. Harko}
\email{harko@hkucc.hku.hk} \affiliation{ Department of Physics,
The University of Hong Kong, Pokfulam Road, Hong Kong, Hong Kong
SAR, P. R. China}

\author{M. K. Mak}
\email{mkmak@vtc.edu.hk} \affiliation{ Department of Physics, The
University of Hong Kong, Pokfulam Road, Hong Kong, Hong Kong SAR,
P. R. China}

\date{\today}

\begin{abstract}
A class of exact solutions of the gravitational field equations in
the vacuum on the brane are obtained by assuming the existence of
a conformal Killing vector field, with non-static and non-central
symmetry. In this case the general solution of the field equations
can be obtained in a parametric form in terms of the Bessel
functions. The behavior of the basic physical parameters
describing the non-local effects generated by the gravitational
field of the bulk (dark radiation and dark pressure) is also
considered in detail, and the equation of state satisfied at
infinity by these quantities is derived. As a physical application
of the obtained solutions we consider the behavior of the angular
velocity of a test particle moving in a stable circular orbit. The
tangential velocity of the particle is a monotonically increasing
function of the radial distance and, in the limit of large values
of the radial coordinate, tends to a constant value, which is
independent on the parameters describing the model. Therefore a
brane geometry admitting a one-parameter group of conformal
motions may provide an explanation for the dynamics of the neutral
hydrogen clouds at large distances from the galactic center, which
is usually explained by postulating the existence of the dark
matter.

\end{abstract}

\pacs{04.50.+h, 04.20.Jb, 04.20.Cv, 95.35.+d}

\maketitle

\section{Introduction}

In two recent papers \cite{Ha03}, \cite{Ma04} several classes of
conformally symmetric solutions of the static gravitational field
equations in the brane world scenario \cite{RS99a}, in which our
Universe is identified to a domain wall in a $5$-dimensional
anti-de Sitter space-time, have been obtained. The solutions have
been derived by using some methods from Lie group theory. As a
group of admissible transformations the one-parameter group of
conformal motions has been considered. More exactly, as a starting
point it has been assumed that the metric tensor $g_{\mu \nu }$ on
the brane has the property $L_{\xi }g_{\mu \nu }=\psi \left(
r\right) g_{\mu \nu }$, where the left-hand side is the Lie
derivative of the metric tensor, describing the gravitational
field in vacuum on the brane, with respect to the vector field
$\xi ^{\mu }$, and $\psi $, the conformal factor, is an arbitrary
function of the radial coordinate $r$.

The assumption that in the brane world model the vacuum outside a
matter distribution has a self-similar structure is not at all
arbitrary; it has a strong theoretical and even observational
basis. The assumption that the space-time admits, beside the
spherical symmetry, a one-parameter group of conformal motions, is
a particular case of the geometric self-similarity, a property of
the metric which has extensively been investigated in the
literature (for a review of this important subject see
\cite{CaCo99}). Geometric similarity should be distinguished from
physical similarity, which is a property of the matter fields.
These two properties are not necessarily equivalent. A "similarity
hypothesis", which assumes that in a variety of physical
situations, solutions of strongly non-linear differential
equations may evolve naturally to a self-similar form, even if
they start out in a more complicated form, has also been proposed
\cite{CaCo00}. The expansion of the Universe from the big bang and
the collapse of a star to a singularity tend to self-similarity in
some circumstances. Spherically symmetric cosmological
fluctuations might naturally evolve from complex initial
conditions via the Einstein equations to a self-similar form.
Self-similar asymptotics can be obtained in a wide range of
contexts in fluid dynamics and for orthogonally transitive $G_2$
cosmological models. The presence of the voids in universe and the
large scale structure can be described by self-similar asymptotic
solutions of the Friedmann equations \cite{To97}. All hypersurface
homogeneous locally rotationally symmetric spacetimes, which admit
conformal symmetries have been determined in \cite{Ap}, and the
symmetry vectors were explicitly obtained. Hence conformal
symmetry is generically possible, it is a mathematical property of
the solutions of many non-linear differential equations and in
some physical/astrophysical phenomena this property has been
effectively observed.

Due to the correction terms coming from the extra dimensions, at
very high energies significant deviations from the Einstein theory
occur in the brane world models \cite{SMS00} (for a recent review
of the brane world theories see \cite{Mar04}). Gravity is largely
modified at the electro-weak scale $1$ TeV. The cosmological
implications of the brane world theories have been extensively
investigated in the physical literature.

For standard general relativistic spherical compact objects the
exterior space-time is described by the Schwarzschild metric. In
the five dimensional brane world models, the high energy
corrections to the energy density, together with the Weyl stresses
from bulk gravitons, imply that on the brane the exterior metric
of a static star is no longer the Schwarzschild metric
\cite{Da00}, \cite{GeMa01}. The presence of the Weyl stresses also
means that the matching conditions do not have a unique solution
on the brane; the knowledge of the five-dimensional Weyl tensor is
needed as a minimum condition for uniqueness. The static vacuum
gravitational field equations on the brane depend on the generally
unknown Weyl stresses, which can be expressed in terms of two
functions, called the dark radiation $U$ and the dark pressure $P$
terms (the projections of the Weyl curvature of the bulk,
generating non-local brane stresses) \cite{Ha03}, \cite{Mar04}.

Generally, the vacuum field equations on the brane can be reduced
to a system of two ordinary differential equations, which describe
all the geometric properties of the vacuum as functions of the
dark pressure and dark radiation terms \cite{Ha03}. In order to
close the system of vacuum field equations on the brane a
functional relation between these two quantities is necessary.
Hence a first possible approach to the study of the vacuum brane
consists in adopting an explicit equation of state for the dark
pressure as a function of the dark radiation. This method has been
systematically considered in \cite{Ha03}, where some classes of
exact solutions of the vacuum gravitational field equations on the
brane have been derived and a homology theorem for spherically
symmetric vacuum on the brane has also been proven.

By using various assumptions and ansatze several classes of exact
vacuum or interior solutions of the gravitational field equations
on the brane have been obtained in \cite {Da00} (the solution has
the mathematical form of the Reissner-Nordstrom solution, in which
a tidal Weyl parameter plays the role of the electric charge of
the general relativistic solution), in \cite{GeMa01} (the exterior
solution also matches a constant density interior) and in
\cite{sol}.

A different approach, which avoids the consideration of some {\it
ad hoc} equations of state for the dark pressure consists in
assuming that the vacuum brane has some particular symmetries, and
to investigate these symmetries by using Lie group techniques
\cite{Ha03}, \cite{Ma04}. As for the vector field $\xi ^{\mu }$
generating the symmetry one can assume that it is static and
spherically symmetric \cite{Ha03} or non-static and spherically
symmetric \cite{Ma04}. In both cases the gravitational field
equations, describing the static vacuum brane with geometric
self-similarity can be integrated in Schwarzschild coordinates,
and several classes of exact solutions, corresponding to a brane
admitting a one-parameter group of motions, can be obtained. The
general solution of the field equations depends on some arbitrary
integration constants. The main advantage of imposing geometric
self-similarity via a group of conformal motions is that this
condition also uniquely fixes the mathematical form of the dark
radiation and dark pressure terms, respectively, which describe
the non-local effects induced by the gravitational field of the
bulk. Thus there is no need to impose an arbitrary relation
between the dark radiation and the dark pressure.

As a possible physical application of the conformally symmetric
solutions with non-static symmetry of the spherically symmetric
gravitational field equations in the vacuum  on the brane the
behavior of the angular velocity $v_{tg}$ of the test particles in
stable circular orbits has been considered in \cite{Ma04}. In this
case the conformal factor $\psi $, together with two constants of
integration, uniquely determines the rotational velocity of the
particle. In the limit of large radial distances and for a
particular set of values of the integration constants the angular
velocity tends to a constant value. This behavior is typical for
massive particles (hydrogen clouds) outside galaxies \cite{Bi87},
and is usually explained by postulating the existence of the dark
matter. Thus, the rotational galactic curves can be naturally
explained in brane world models without introducing any additional
hypothesis. The galaxy is embedded in a modified, spherically
symmetric geometry, generated by the non-zero contribution of the
Weyl tensor from the bulk. The extra-terms, which can be described
in terms of the dark radiation term $U$ and the dark pressure term
$P$, act as a "matter" distribution outside the galaxy. The
existence of the dark radiation term generates an equivalent mass
term $M_U$, which is linearly increasing with the distance, and
proportional to the baryonic mass of the galaxy, $M_{U}(r)\approx
M_{B}(r/r_{0})$. The particles moving in this geometry feel the
gravitational effects of $U$, which can be expressed in terms of
an equivalent mass. Moreover, the limiting value $v_{tg\infty }$
of the angular velocity can be obtained as function of the total
baryonic mass $M_B$ and radius $r_0$ of the galaxy as $v_{tg\infty
}\approx (2/\sqrt{3})\sqrt{GM_{B}/r_{0}}+ (1/12\sqrt{3})\left(
GM_{B}/r_{0}\right) ^{3/2}$ \cite{Ma04}.  For a galaxy with
baryonic mass of the order $10^{9}M_{\odot }$ and radius of the
order of $r_{0}\approx 70$ kpc, we have $v_{tg\infty }\approx 287$
km/s, which is of the same order of magnitude as the observed
value of the angular velocity of the galactic rotation curves. In
the framework of this model all the relevant physical parameters
(metric tensor components, dark radiation and dark pressure terms)
can be obtained as functions of the tangential velocity, and hence
they can be determined observationally.

It is the purpose of the present paper to consider vacuum
spacetimes on the brane admitting a one parameter group of
conformal motions, with the vector field generating the motion
being non-static and having a more general, non-central symmetry.
More exactly, we assume that the vector field ${\bf \xi }$
generating the conformal symmetry has the general non-static and
non-spherically symmetric form ${\bf \xi }=\xi ^{0}\left(
t,r\right)
\partial /\partial t+\xi ^{1}\left( t,r\right)
\partial /\partial r+\xi ^{2}\left( \theta ,\phi \right)
\partial /\partial \theta +\xi ^{3}\left( \theta ,\phi \right)
\partial /\partial \phi $, that is, we also introduce an explicit
dependence of ${\bf \xi }$ on the angular coordinates $\theta $
and $\phi $. With this assumption the metric tensor components can
be obtained as functions of the conformal factor $\psi $ and of
the radial coordinate $r$. The gravitational field equations
describing the conformally symmetric vacuum brane with general
symmetry can be reduced to a single differential-integral
equation, whose solution can be obtained in terms of the modified
Bessel functions. The expressions for the metric coefficients, the
dark energy and the dark pressure can be expressed in an exact
parametric form.

As a physical application of the derived solutions we consider the
behavior of a test particle in a stable circular orbit on the
conformally symmetric brane. Similar to the spherically symmetric
non-static case, the angular velocity is always a monotonically
increasing function of the radial distance and tends, in the limit
of large distances, to a constant value. This behavior strongly
suggest the possibility that a self similar geometry on the brane,
generated by a general, non-static and non-centrally symmetric
group of conformal motions, may provide an explanation for the
dynamics of the neutral hydrogen clouds at large distances from
the galactic center, which is usually explained by postulating the
existence of the dark matter.

The present paper is organized as follows. The basic equations
describing the spherically symmetric gravitational field equations
in the vacuum on the static brane are derived in Section II. The
general form of the metric tensor and the field equations for
vacuum brane space-times admitting a one parameter group of
conformal motions, with non-static and non-central conformal
symmetry, are derived in Section III. In Section IV the general
solution of the gravitational field equations is obtained. The
behavior of the angular velocity of a test particle in stable
circular motion is considered in Section V. We conclude and
discuss our results in Section VI.

\section{The field equations for a static, spherically symmetric vacuum brane}

We start by considering a $5$-dimensional space-time (the bulk),
with a single $4$-dimensional brane, on which gravity is confined.
The $4-$D brane
world $\left( ^{(4)}M,g_{\mu \nu }\right) $ is located at a hypersurface $%
\left( B\left( X^{A}\right) =0\right) $ in the $5-$D bulk
space-time $\left(
^{(5)}M,g_{AB}\right) $, of which coordinates are described by $%
X^{A},A=0,1,...,4$. The action of the system is given by
\cite{SMS00}
\begin{equation}
S=S_{bulk}+S_{brane},
\end{equation}
where
\begin{equation}
S_{bulk}=\int_{^{(5)}M}\sqrt{-^{(5)}g}\left[ \frac{1}{2k_{5}^{2}}%
^{(5)}R+^{(5)}L_{m}+\Lambda _{5}\right] d^{5}X,
\end{equation}
and
\begin{equation}
S_{brane}=\int_{^{(4)}M}\sqrt{-^{(5)}g}\left[
\frac{1}{k_{5}^{2}}K^{\pm }+L_{brane}\left( g_{\alpha \beta },\psi
\right) +\lambda _{b}\right] d^{4}x,
\end{equation}
where $k_{5}^{2}=8\pi G_{5}$ is the $5$-D gravitational constant,
$^{(5)}R$ and $^{(5)}L_{m}$ are the $5-$D scalar curvature and
matter Lagrangian in the bulk, respectively. $x^{\mu },\mu
=0,1,2,3$ are the induced $4-$D coordinates of the brane, $K^{\pm
}$ is the trace of the extrinsic curvature on either side of the
brane and $L_{brane}\left( g_{\alpha \beta },\psi \right) $ is the
$4-$D Lagrangian, which is given by a generic functional of the
brane metric $g_{\alpha \beta }$ and of the matter fields $\psi $.
In the following capital Latin indices run in the range $0,...,4$,
while Greek indices take the values $0,...,3$. $\Lambda _{5}$ and
$\Lambda $ are the negative vacuum energy densities in the bulk
and in the brane, respectively.

The Einstein field equations in the bulk are given by
\begin{equation}
^{(5)}G_{IJ}=k_{5}^{2}{\rm  }^{(5)}T_{IJ},\qquad ^{(5)}T_{IJ}=-\Lambda _{5}%
{\rm }^{(5)}g_{IJ}+\delta (B)\left[ -\lambda _{b}{\rm }%
^{(5)}g_{IJ}+T_{IJ}\right] ,
\end{equation}
where
\begin{equation}
^{(5)}T_{IJ}\equiv -2\frac{\delta ^{(5)}L_{m}}{\delta ^{(5)}g^{IJ}}%
+^{(5)}g_{IJ}{\rm }^{(5)}L_{m},
\end{equation}
is the energy-momentum tensor of bulk matter fields, while $T_{\mu
\nu }$ is the energy-momentum tensor localized on the brane and
which is defined by
\begin{equation}
T_{\mu \nu }\equiv -2\frac{\delta L_{brane}}{\delta g^{\mu \nu
}}+g_{\mu \nu }{\rm }L_{brane}.
\end{equation}

The $\delta \left( B\right) $ denotes the localization of brane
contribution. In the $5$-D space-time a brane is a fixed point of
the $Z_{2}$ symmetry. The basic equations on the brane are
obtained by projection of the variables onto the brane world,
because we assume that the gravity on the
brane is confined. The induced $4-$D metric is $%
g_{IJ}=^{(5)}g_{IJ}-n_{I}n_{J}$, where $n_{I}$ is the space-like
unit vector field normal to the brane hypersurface $^{(4)}M$. In
the following we assume that all the matter fields, except
gravitation, are confined to the brane. This implies that
$^{(5)}L_{m}=0$.

Assuming a metric of the form
$ds^{2}=(n_{I}n_{J}+g_{IJ})dx^{I}dx^{J}$, with $n_{I}dx^{I}=d\chi
$ the unit normal to the $\chi =$constant hypersurfaces and
$g_{IJ}$ the induced metric on $\chi =$constant hypersurfaces, the
effective four-dimensional gravitational equations on the brane
(the Gauss equation), take the form \cite{SMS00}:
\begin{equation}
G_{\mu \nu }=-\Lambda g_{\mu \nu }+k_{4}^{2}T_{\mu \nu
}+k_{5}^{4}S_{\mu \nu }-E_{\mu \nu },  \label{Ein}
\end{equation}
where $S_{\mu \nu }$ is the local quadratic energy-momentum
correction
\begin{equation}
S_{\mu \nu }=\frac{1}{12}TT_{\mu \nu }-\frac{1}{4}T_{\mu
}{}^{\alpha }T_{\nu \alpha }+\frac{1}{24}g_{\mu \nu }\left(
3T^{\alpha \beta }T_{\alpha \beta }-T^{2}\right) ,
\end{equation}
and $E_{\mu \nu }$ is the non-local effect from the free bulk
gravitational
field, the transmitted projection of the bulk Weyl tensor $C_{IAJB}$, $%
E_{IJ}=C_{IAJB}n^{A}n^{B}$, with the property $E_{IJ}\rightarrow
E_{\mu \nu }\delta _{I}^{\mu }\delta _{J}^{\nu }\quad $as$\quad
\chi \rightarrow 0$. We have also denoted $k_{4}^{2}=8\pi G$, with
$G$ the usual four-dimensional gravitational constant.

The four-dimensional cosmological constant, $\Lambda $, and the
four-dimensional coupling constant, $k_{4}$, are given by $\Lambda
=k_{5}^{2}\left( \Lambda _{5}+k_{5}^{2}\lambda _{b}^{2}/6\right) /2$ and $%
k_{4}^{2}=k_{5}^{4}\lambda _{b}/6$, respectively. In the limit
$\lambda _{b}^{-1}\rightarrow 0$ we recover standard general
relativity.

The Einstein equation in the bulk and the Codazzi equation also
imply the
conservation of the energy-momentum tensor of the matter on the brane, $%
D_{\nu }T_{\mu }{}^{\nu }=0$, where $D_{\nu }$ denotes the brane
covariant derivative. Moreover, from the contracted Bianchi
identities on the brane it follows that the projected Weyl tensor
should obey the constraint $D_{\nu }E_{\mu }{}^{\nu
}=k_{5}^{4}D_{\nu }S_{\mu }{}^{\nu }$.

The symmetry properties of $E_{\mu \nu }$ imply that in general we
can decompose it irreducibly with respect to a chosen $4$-velocity
field $u^{\mu }$ as \cite{Mar04}
\begin{equation}
E_{\mu \nu }=-k^{4}\left[ U\left( u_{\mu }u_{\nu
}+\frac{1}{3}h_{\mu \nu }\right) +P_{\mu \nu }+2Q_{(\mu }u_{\nu
)}\right] ,  \label{WT}
\end{equation}
where $k=k_{5}/k_{4}$, $h_{\mu \nu }=g_{\mu \nu }+u_{\mu }u_{\nu
}$ projects orthogonal to $u^{\mu }$, the ''dark radiation'' term
$U=-k^{4}E_{\mu \nu }u^{\mu }u^{\nu }$ is a scalar, $Q_{\mu
}=k^{4}h_{\mu }^{\alpha }E_{\alpha
\beta }$ a spatial vector and $P_{\mu \nu }=-k^{4}\left[ h_{(\mu }{\rm }%
^{\alpha }h_{\nu )}{\rm }^{\beta }-\frac{1}{3}h_{\mu \nu }h^{\alpha \beta }%
\right] E_{\alpha \beta }$ a spatial, symmetric and trace-free
tensor.

In the case of the vacuum state we have $\rho =p=0$, $T_{\mu \nu
}\equiv 0$ and consequently $S_{\mu \nu }=0$. Therefore the field
equations describing a static brane take the form
\begin{equation}
R_{\mu \nu }=-E_{\mu \nu }+4\Lambda ,
\end{equation}
with the trace $R$ of the Ricci tensor $R_{\mu \nu }$ satisfying
the condition $R=R_{\mu }^{\mu }=E_{\mu }^{\mu }=0$.

In the vacuum case $E_{\mu \nu }$ satisfies the constraint $D_{\nu
}E_{\mu
}{}^{\nu }=0$. In an inertial frame at any point on the brane we have $%
u^{\mu }=\delta _{0}^{\mu }$ and $h_{\mu \nu }=$diag$\left( 0,1,1,1\right) $%
. In a static vacuum $Q_{\mu }=0$ and the constraint for $E_{\mu
\nu }$ takes the form \cite{GeMa01}
\begin{equation}
\frac{1}{3}D_{\mu }U+\frac{4}{3}UA_{\mu }+D^{\nu }P_{\mu \nu
}+A^{\nu }P_{\mu \nu }=0,
\end{equation}
where $D_{\mu }$ is the projection (orthogonal to $u^{\mu }$) of
the covariant derivative and $A_{\mu }=u^{\nu }D_{\nu }u_{\mu }$
is the 4-acceleration. In the static spherically symmetric case we may chose $%
A_{\mu }=A(r)r_{\mu }$ and $P_{\mu \nu }=P(r)\left( r_{\mu }r_{\nu }-\frac{1%
}{3}h_{\mu \nu }\right) $, where $A(r)$ and $P(r)$ (the ''dark
pressure'') are some scalar functions of the radial distance $r$,
and $r_{\mu }$ is a unit radial vector \cite{Da00}.

We chose the static spherically symmetric metric on the brane in
the standard form
\begin{equation}
ds^{2}=-e^{\nu \left( r\right) }dt^{2}+e^{\lambda \left( r\right)
}dr^{2}+r^{2}\left( d\theta ^{2}+\sin ^{2}\theta d\phi ^{2}\right)
. \label{line}
\end{equation}

Then the gravitational field equations and the effective
energy-momentum tensor conservation equation in the vacuum take
the form \cite{Ha03}, \cite{Ma04}
\begin{equation}
-e^{-\lambda }\left( \frac{1}{r^{2}}-\frac{\lambda ^{\prime }}{r}\right) +%
\frac{1}{r^{2}}=\frac{48\pi G}{k^{4}\lambda _{b}}U+\Lambda,
\label{f1}
\end{equation}
\begin{equation}
e^{-\lambda }\left( \frac{\nu ^{\prime }}{r}+\frac{1}{r^{2}}\right) -\frac{1%
}{r^{2}}=\frac{16\pi G}{k^{4}\lambda _{b}}\left( U+2P\right)
-\Lambda, \label{f2}
\end{equation}
\begin{equation}
e^{-\lambda }\left( \nu ^{\prime \prime }+\frac{\nu ^{\prime 2}}{2}+\frac{%
\nu ^{\prime }-\lambda ^{\prime }}{r}-\frac{\nu ^{\prime }\lambda ^{\prime }%
}{2}\right) =\frac{32\pi G}{k^{4}\lambda _{b}}\left(
U-P\right)-2\Lambda  , \label{f3}
\end{equation}
\begin{equation}
\nu ^{\prime }=-\frac{U^{\prime }+2P^{\prime
}}{2U+P}-\frac{6P}{r\left( 2U+P\right) },  \label{f4}
\end{equation}
where we denoted $^{\prime }=d/dr$. In the following we denote
$\beta =16\pi G/k^{4}\lambda _{b}$.

\section{Field equations on a vacuum brane with non-static and non-central conformal symmetry}

The system of the field equations for the vacuum on the brane is
under-determined. A functional relation between the dark energy
$U$ and the dark pressure $P$ must be specified in order to solve
the equations. An alternative method, which avoids {\it ad hoc }
specifications, is to assume that the brane is mapped conformally
onto itself along the direction $\xi $, so that
\begin{equation}
L_{\xi }g_{\mu \nu }=g_{\mu \nu ,\lambda }\xi ^{\lambda
}+g_{\lambda \nu }\xi _{,\mu }^{\lambda }+g_{\mu \lambda }\xi
_{,\nu }^{\lambda }=\psi g_{\mu \nu },  \label{1}
\end{equation}
where $\psi $ is the conformal factor. In the following we shall
assume that $\psi $ is a function of the radial coordinate only,
$\psi =\psi (r)$. As for the choice of $\xi $, in the framework of
the standard general relativity, Herrera and Ponce de Leon
\cite{He} assumed that the vector field generating the conformal
symmetry is static and spherically symmetric,
\begin{equation}
{\bf \xi }=\xi ^{0}\left( r\right) \frac{\partial }{\partial
t}+\xi ^{1}\left( r\right) \frac{\partial }{\partial r}.
\label{2}
\end{equation}

Using this form of the conformal vector in Eqs. (\ref{1}) one
obtains
\begin{equation}
\xi^{0}=A, \xi ^{1}=\frac{B}{2} r\exp \left( -\frac{\lambda
}{2}\right),
\end{equation}
and
\begin{equation}
\psi \left( r\right) =B\exp \left( -\frac{\lambda }{2}\right),
\exp \left( \nu \right) =C^{2}r^{2},
\end{equation}
respectively, where $A$, $B$, $C$ are constants. $A$ may be set to
zero since $A\partial /\partial t$ is a Killing vector and $B$ may
be set to $1$ by a rescaling $\xi \rightarrow B^{-1}\xi $, $\psi
\rightarrow B^{-1}\psi $, which leaves Eqs. (\ref{1}) invariant.

This form of $\xi $ gives the most general $\xi $ invariant under
the Killing symmetries, that is $\left[ \partial /\partial t,{\bf
\xi }\right] =0=\left[ {\bf X}_{\alpha },{\bf \xi }\right] $,
where ${\bf X}_{\alpha }$ generates $SO\left( 3\right) $. The
corresponding form of the metric, obtained by imposing static
conformal symmetry, has been used to study the properties of
compact anisotropic general relativistic objects \cite{He} and to
investigate the properties of charged strange stars \cite{MaD}.
The general solution of the vacuum brane gravitational field
equations for this choice of ${\bf \xi } $ has been obtained in
\cite{Ha03}.

A more general conformal symmetry has been proposed by Maartens
and
Maharajah \cite{MaMa90}, which generalizes the isotropic conformal vector $%
t\partial /\partial t+r\partial /\partial r$ of the Minkowski
space-time, but weakens the static symmetry of $\xi $ in Eq.
(\ref{2}):
\begin{equation}
{\bf \xi }=\xi ^{0}\left( t,r\right) \frac{\partial }{\partial
t}+\xi ^{1}\left( t,r\right) \frac{\partial }{\partial r}.
\label{3}
\end{equation}

With this choice the conformal vector is non-static and
spherically symmetric. With the assumption that the conformal
factor $\psi $ is static, $\psi =\psi \left( r\right) $ and with
this form of ${\bf \xi }$ Eqs. (\ref{1}) give \cite{MaMa90}:
\begin{equation}
\xi ^{0}=A+\frac{1}{2}\frac{k}{B}t,
\end{equation}
\begin{equation}\label{nu}
e^{\nu }=C^{2}r^{2}\exp \left( -2kB^{-1}\int \frac{dr}{r\psi
}\right) ,
\end{equation}
\begin{equation}\label{l}
 \psi =Be^{-\lambda /2},
\end{equation}
where $k$ is a separation constant and $A$, $B$ and $C$ are
integration constants. Without any loss of generality we can chose
$A=0$. Thus in this model we obtain for the vector field ${\bf \xi
}$ the expression
\begin{equation}\label{vec}
{\bf \xi =}\frac{1}{2}\frac{k}{B}t\frac{\partial }{\partial
t}+\frac{r\psi \left( r\right) }{2}\frac{\partial }{\partial r}.
\end{equation}

In the framework of the standard general relativity the static
solutions of the gravitational field equations with metric tensor
components given by Eqs. (\ref{nu}) and (\ref{l}) have been
obtained, for anisotropic and charged compact objects, by Maartens
and Maharaj \cite{MaMa90}. All the solutions of the vacuum
gravitational field equations on the brane having the conformal
symmetry generated by the vector $\bf \xi $ given by Eq.
(\ref{vec}) have been obtained in \cite{Ma04}.

In the following we consider a generalization of the conformal
vector $\bf \xi $ by relaxing the condition of the spherical
symmetry. Hence we assume a more general form of the vector field
${\bf \xi }$, generating the conformal symmetry, namely
\begin{equation}
{\bf \xi }=\xi ^{0}\left( t,r\right) \frac{\partial }{\partial
t}+\xi ^{1}\left( t,r\right) \frac{\partial }{\partial r}+\xi
^{2}\left( \theta ,\phi \right) \frac{\partial }{\partial \theta
}+\xi ^{3}\left( \theta ,\phi \right) \frac{\partial }{\partial
\phi }.  \label{csi}
\end{equation}

We still assume that the conformal factor is spherically
symmetric, that is, we take $\psi =\psi \left( r\right)$. With the
use of this form of ${\bf \xi }$, Eqs. (\ref{1}) give
\begin{equation}\label{dcsi1}
\nu ^{\prime }\xi ^{1}+2\frac{\partial \xi ^{0}}{\partial t}=\psi
\left( r\right) ,
\end{equation}
\begin{equation}\label{dcsi2}
\lambda ^{\prime }\xi ^{1}+2\frac{\partial \xi ^{1}}{\partial r}%
=\psi \left( r\right) ,
\end{equation}
\begin{equation}\label{dcsi3}
\frac{\xi ^{1}}{r}+\frac{\partial \xi ^{2}}{\partial \theta
}=\frac{\psi \left( r\right) }{2},
\end{equation}
\begin{equation}\label{dcsi4}
\frac{\xi ^{1}}{r}+\frac{\cos \theta }{\sin \theta }\xi
^{2}+\frac{\partial \xi ^{3}}{\partial \phi }=\frac{\psi \left( r\right) }{2}%
.
\end{equation}

From Eq. (\ref{dcsi3}) it follows that
\begin{equation}
\xi ^{1}\left( r\right) =\frac{r\left( \psi -\alpha \right) }{2},
\end{equation}
\begin{equation}
\xi ^{2}\left( \theta ,\phi \right) =\frac{\alpha }{2}\theta
+\frac{dF\left( \phi \right) }{d\phi },
\end{equation}
where $\alpha $ is an arbitrary separation constant and $F$ an
arbirary function of the azimuthal angle $\phi $. Then Eq. (\ref
{dcsi4}) gives
\begin{equation}
\xi ^{3}\left( \theta ,\phi \right) =\frac{\alpha }{2}\left(
1-\theta \cot \theta \right) \phi -\cot \theta F\left( \phi
\right) +G\left( \theta \right) ,
\end{equation}
where $G\left( \theta \right) $ is an arbitrary integration
function. From  Eq. (\ref{dcsi1}) we obtain
\begin{equation}
\xi ^{0}\left( t\right) =\frac{k}{2}t+A,
\end{equation}
where $k$ is a separation constant and $A$ is an integration
constant, which can be taken to be zero without any loss of
generality, $A=0$. Hence the metric tensor component $\exp \left(
\nu \right) $ is given by
\begin{equation}
\exp \left( \nu \right) =C^{2}r^{2}\exp \left[ 2\left( \alpha
-k\right) \int \frac{dr}{r\left( \psi -\alpha \right) }\right] ,
\label{nuu}
\end{equation}
where $C$ is an arbitrary integration constant. Finally, the
metric tensor component $\exp \left( \lambda \right) $ can be
obtained in the form
\begin{equation}
\exp \left( \lambda \right) =\frac{B^{2}}{\left( \psi -\alpha \right) ^{2}}%
\exp \left[ 2\alpha \int \frac{dr}{r\left( \psi -\alpha \right)
}\right] , \label{lam}
\end{equation}
where $B$ is an arbitrary constant of integration.

Therefore the requirement of the conformal symmetry on the brane
completely fixes the form of the metric tensor, as a function of
the conformal factor $\psi $. As for the vector field ${\bf \xi
}$, it is given by
\begin{equation}
{\bf \xi =}\frac{k}{2}t\frac{\partial }{\partial t}+\frac{r\left(
\psi
-\alpha \right) }{2}\frac{\partial }{\partial r}+\left[ \frac{\alpha }{2}%
\theta +\frac{dF\left( \phi \right) }{d\phi }\right] \frac{\partial }{%
\partial \theta }+\left[ \frac{\alpha }{2}\left( 1-\theta \cot \theta
\right) \phi -\cot \theta F\left( \phi \right) +G\left( \theta \right) %
\right] \frac{\partial }{\partial \phi }.
\end{equation}

With the use of the representations given by Eqs. (\ref{nuu}) and (\ref{lam}%
) for the metric tensor components, the gravitational field
equations (\ref {f1})-(\ref{f3}) on the vacuum brane with general,
non-central conformal symmetry, take the form
\begin{equation}
-\frac{\left( \psi -\alpha \right) ^{2}}{B^{2}}e^{-2\alpha \int \frac{dr}{%
r\left( \psi -\alpha \right) }}\left[ \frac{2\psi ^{\prime
}}{r\left( \psi
-\alpha \right) }-\frac{2\alpha }{r^{2}\left( \psi -\alpha \right) }+\frac{1%
}{r^{2}}\right] +\frac{1}{r^{2}}=3\beta U+\Lambda ,  \label{br1}
\end{equation}
\begin{equation}
\frac{\left( \psi -\alpha \right) ^{2}}{B^{2}}e^{-2\alpha \int \frac{dr}{%
r\left( \psi -\alpha \right) }}\left[ \frac{2\left( \alpha -k\right) }{%
r^{2}\left( \psi -\alpha \right) }+\frac{3}{r^{2}}\right] -\frac{1}{r^{2}}%
=\beta \left( U+2P\right)-\Lambda ,  \label{br2}
\end{equation}
\begin{equation}
\frac{\left( \psi -\alpha \right) ^{2}}{B^{2}}e^{-2\alpha \int \frac{dr}{%
r\left( \psi -\alpha \right) }}\left[ \frac{2\psi ^{\prime
}}{r\left( \psi -\alpha \right) }-\frac{2k}{r^{2}\left( \psi
-\alpha \right) }-\frac{k\left(
\alpha -k\right) }{r^{2}\left( \psi -\alpha \right) ^{2}}+\frac{1}{r^{2}}%
\right] =\beta \left( U-P\right)-\Lambda .  \label{br3}
\end{equation}

\section{General solution of the gravitational field equations for the
vacuum brane with non-central conformal symmetry}

By multiplying Eq. (\ref{br3}) with $2$, adding the obtained
equation to Eq. (\ref{br2}) and equating the resulting equation
with Eq. (\ref{br1}) gives the following differential-integral
equation, satisfied by the conformal factor $\psi \left( r\right)
$:
\begin{equation}
3r\left( \psi -\alpha \right) \psi ^{\prime }+3\psi ^{2}-3\left(
k+2\alpha
\right) \psi +3\alpha ^{2}+2\alpha k+k^{2}-B^{2}e^{2\alpha \int \frac{dr}{%
r\left( \psi -\alpha \right) }}-4\Lambda B^2e^{2\alpha \int \frac{dr}{%
r\left( \psi -\alpha \right) }}r^2=0.  \label{fin}
\end{equation}

In order to solve Eq. (\ref{fin}) we introduce a new variable $u$
defined as
\begin{equation}
\ln u=\alpha \int \frac{dr}{r\left( \psi -\alpha \right) }.
\end{equation}

In terms of $u$ Eq. (\ref{fin}) can be expressed in the form
\begin{equation}
-3\frac{\alpha ^{2}}{r}u^{2}\frac{u^{\prime \prime }}{u^{\prime
3}}+3\alpha \left( \alpha -k\right) \frac{1}{r}\frac{u}{u^{\prime
}}+k^{2}-k\alpha -B^{2}u^{2}-4\Lambda B^2u^2r^2=0.  \label{fin1}
\end{equation}

With the use of the identities $1/u^{\prime }=dr/du$ and
$u^{\prime \prime }/u^{\prime 3}=-d^{2}r/du^{2}$, Eq. (\ref{fin1})
can be transformed into the following differential equation:
\begin{equation}
u^{2}\frac{d^{2}r}{du^{2}}-mu\frac{dr}{du}-\left(
l^{2}u^{2}-n^{2}\right) r-12l\Lambda u^2r^3=0,  \label{fin2}
\end{equation}
where we denoted
\begin{equation}
m=\frac{k-\alpha }{\alpha },l^{2}=\frac{B^{2}}{3\alpha ^{2}},n^{2}=\frac{%
k\left( k-\alpha \right) }{3\alpha ^{2}}.
\end{equation}

In the following we assume that we can neglect the nonlinear term
in Eq. (\ref{fin2}). This approximation is valid in the case of
very small numerical values of the parameters $l$ and $\Lambda $
and for small values of $r$. Therefore the basic equation
describing the geometry of the vacuum brane with geometric
self-similarity with the effect of the cosmological constant
ignored is given by the following linear, Bessel type differential
equation:
\begin{equation}
u^{2}\frac{d^{2}r}{du^{2}}-mu\frac{dr}{du}-\left(
l^{2}u^{2}-n^{2}\right) r=0,  \label{fin3}
\end{equation}

The general solution of Eq. (\ref{fin3}) is given by
\begin{equation}\label{r}
r\left( u\right) =u^{s}\left[ C_{1}I_{p}\left( lu\right)
+C_{2}K_{p}\left( lu\right) \right],
\end{equation}
where $C_{1}$ and $C_{2}$ are arbitrary constants of integration,
$s=\left( 1+m\right) /2=k/2\alpha $, $p=\sqrt{\left( 1+m\right)
^{2}-4n^{2}}/2=\sqrt{s(2-s)/3}$, and $I_{p}\left( x\right) $ and
$K_{p}\left( x\right) $ are the modified Bessel functions of the
first and second kind, respectively, satisfying the differential
equation $x^{2}y^{\prime \prime }+xy^{\prime }-\left(
x^{2}+p^{2}\right) y=0$ \cite{AbSt}. For all values of $p$,
$I_p(x)$ and $K_p(x)$ are linearly independent. $I_p(x)$ and
$K_p(x)$ are real and positive when $p>-1$ and $x>0$. Since by
definition $l>0$, it follows that the parameter $u$ must be
non-negative, $u\geq 0$. On the other hand the condition that $p$
must be a real number, $p\in R$, imposes the constraint $0<s\leq
2$ on the numerical parameter $s$.

To obtain a correct physical description, the function $r(u)$ must
be an increasing function of the variable $u$ and obey the
conditions $\lim_{u\rightarrow 0}r(u)=0$ and $\lim_{u\rightarrow
\infty }r(u)=\infty $, respectively. For small values of the
arguments and for fixed $p$ the modified Bessel functions behave
like $I_{p}(x)\sim \left( x/2\right) ^{p}/\Gamma (p+1)$ and
$K_{p}(x)\sim \Gamma \left( p\right) \left( x/2\right) ^{-p}/2$,
respectively, where $\Gamma $ is the Euler gamma function, defined as $%
\Gamma \left( z\right) =\int_{0}^{\infty }t^{z-1}e^{-z}dt$
\cite{AbSt}. Hence in the limit of small $u$ we obtain
$\lim_{u\rightarrow 0}r(u)\sim u^{s}\left[ C_{1}\left( l/2\right)
^{p}u^{p}/\Gamma (p+1)+C_{2}\Gamma \left( p\right) \left(
l/2\right) ^{-p}u^{-p}/2\right] $. In order to obtain $%
\lim_{u\rightarrow 0}r(u)=0$, the condition $s>p$ must necessarily
hold. This condition is satisfied for values of $s$ so that
$s>1/2$.

In the opposite limit of large values of the argument, the
modified Bessel functions $I_{p}(x)$ and $K_{p}(x)$ have the
asymptotic behaviors $I_{p}(x)\sim \exp \left( x\right) /\sqrt{2\pi x}$ and $%
K_{p}(x)\sim \sqrt{\pi /2x}\exp \left( -x\right) $, respectively \cite{AbSt}. Hence $%
\lim_{u\rightarrow \infty }r(u)=\lim_{u\rightarrow \infty }\left[
\left( C_{1}/\sqrt{2\pi l}\right) u^{s-1/2}\exp \left( lu\right)
\right] =\infty $. To obtain the correct form of the limit it is
also necessary that $C_{1}>0$. The numerical values of the
parameter $s$ are restricted to the range $s\in (1/2,2]$.

 Therefore the general solution of the gravitational field equations on the
vacuum brane with non-static and non-central conformal symmetry
can be expressed in an exact parametric form, with $u$ taken as
parameter, and given by
\begin{equation}
e^{\nu }=C^{2}u^{2(1-s)}\left[ C_{1}I_{p}\left( lu\right)
+C_{2}K_{p}\left( lu\right) \right] ^{2},
\end{equation}
\begin{equation}\label{lambda}
e^{\lambda }=\frac{3l^{2}u^{2s}\left[ C_{1}I_{p}\left( lu\right)
+C_{2}K_{p}\left( lu\right) \right] ^{2}}{\left( \frac{dr}{du}%
\right) ^{2}},
\end{equation}
\begin{equation}
3\beta U=\frac{1}{3r^{2}}\left\{ 1+\frac{1}{l^{2}}\left[ \left( \frac{d}{du}%
\ln r\right) ^{2}+\frac{2(2s-1)}{u}\left( \frac{d}{du}\ln r\right) +\frac{%
2n^{2}}{u^{2}}\right] \right\} ,
\end{equation}
\begin{equation}
3\beta P=\frac{1}{3l^{2}r^{2}}\left[ 4\left( \frac{d}{du}\ln r\right) ^{2}-%
\frac{2(2s-1)}{u}\left( \frac{d}{du}\ln r\right) +l^{2}-\frac{n^{2}}{u^{2}}%
\right] -\frac{2}{r^{2}}.
\end{equation}

The variation of the metric coefficient exp$\left(\nu
\right)/C^2C_2^2$ is represented, as a function of $r/C_2$, in
Fig. 1

\vspace{0.2in}
\begin{figure}[h]
\includegraphics{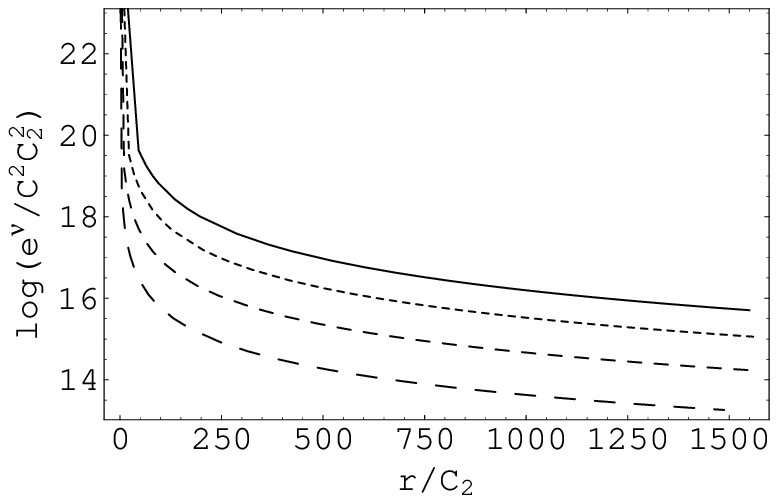}
\caption{Variation, as a function of the radial distance $r/C_2$,
of the metric coefficient exp$\left(\nu \right)/C^2C_2^2$ (in a
logarithmic scale) for a static, conformally symmetric vacuum
space-time on the brane, for $l=10^{-9}$, $R_0=C_1/C_2=10^{-5}$
and different values of $s$: $s=1.1$ (solid curve), $s=1.2$
(dotted curve), $s=1.3$ (dashed curve) and $s=1.4$ (long dashed
curve). } \label{FIG1}
\end{figure}

For the chosen values of the parameters $\exp(\nu )$ is a
decreasing function of the radial distance $r$. The variation of
the metric function $\exp{(\lambda )}/3l^2$ is represented in Fig.
2.

\vspace{0.2in}
\begin{figure}[h]
\includegraphics{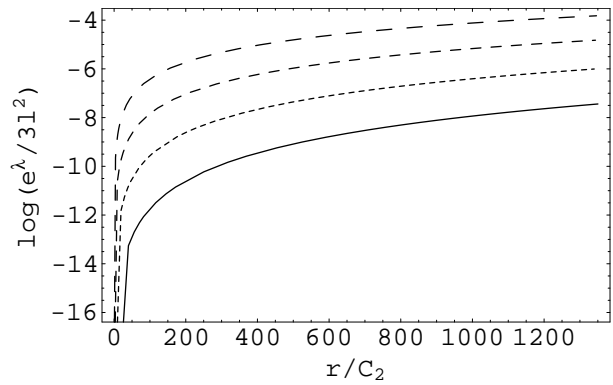}
\caption{Variation, as a function of the radial distance $r/C_2$,
of the metric coefficient exp$\lambda /3l^2$ (in a logarithmic
scale) for a static, conformally symmetric vacuum space-time on
the brane, for $l=10^{-9}$, $R_0=C_1/C_2=10^{-5}$ and different
values of $s$: $s=1.1$ (solid curve), $s=1.2$ (dotted curve),
$s=1.3$ (dashed curve) and $s=1.4$ (long dashed curve). }
\label{FIG2}
\end{figure}

The metric function $\exp(\lambda )$ is a monotonically increasing
function of the coordinate $r$. In the limit of large values of
$r$, $\exp(\lambda )$ tends to a constant value. The variation of
the dark radiation term $3\beta C_2^2U$ is represented, as a
function of $r/C_2$, in Fig. 3.

\vspace{0.2in}
\begin{figure}[h]
\includegraphics{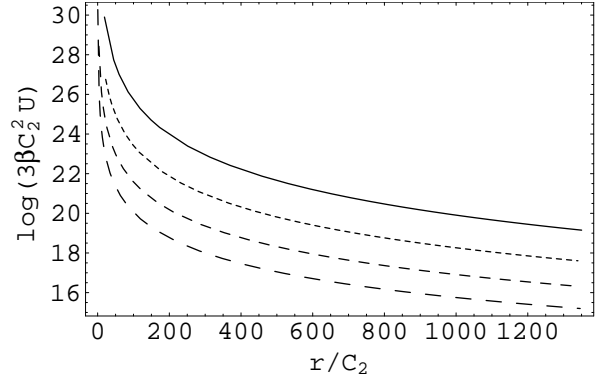}
\caption{Variation, as a function of the radial distance $r/C_2$,
of the dark radiation term $3\beta C_{2}^{2}U$ (in a logarithmic
scale) for a static, conformally symmetric vacuum space-time on
the brane, for $l=10^{-9}$, $R_0=C_1/C_2=10^{-3}$ and different
values of $s$: $s=1.1$ (solid curve), $s=1.2$ (dotted curve),
$s=1.3$ (dashed curve) and $s=1.4$ (long dashed curve). }
\label{FIG3}
\end{figure}

The dark radiation term is positive for all values of the radial
coordinate $r$, $U(r)\geq 0$, $\forall r\in \left( 0,\infty
\right) $. In the limit of large $r$, $U$ tends to zero,
$\lim_{r\rightarrow \infty }U(r)=0$. The variation of the absolute
value of the dark pressure $3\beta C_{2}^{2}\left| P\right| $ as a
function of $r/C_2$ is represented in Fig. 4.

\vspace{0.2in}
\begin{figure}[h]
\includegraphics{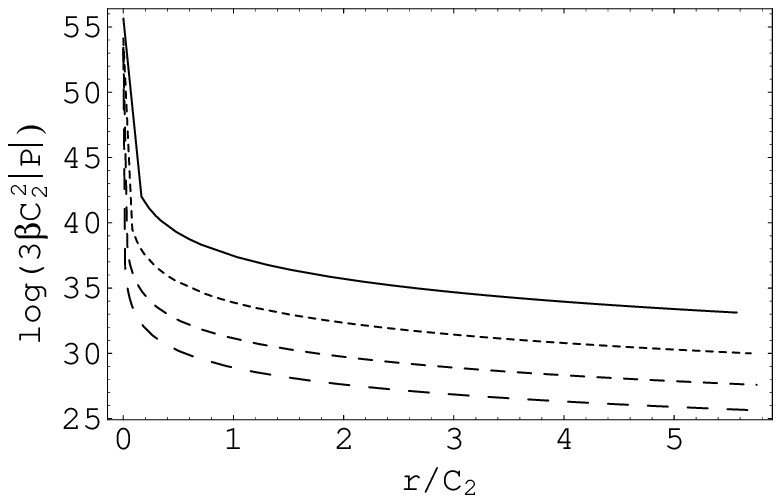}
\caption{Variation, as a function of the radial distance $r/C_2$,
of the dark pressure term $3\beta C_{2}^{2}\left| P\right| $ for a
static, conformally symmetric vacuum space-time on the brane, for
$l=10^{-9}$, $R_0=C_1/C_2=10^{-3}$ and different values of $s$:
$s=1.1$ (solid curve), $s=1.2$ (dotted curve), $s=1.3$ (dashed
curve) and $s=1.4$ (long dashed curve). } \label{FIG4}
\end{figure}

In the present model the dark pressure is negative, satisfying the
condition $P(r)\leq 0$, $\forall r\in \left( 0,\infty \right) $.
In the large time limit, similar to the dark radiation term, the
dark pressure also tends to zero, $\lim_{r\rightarrow \infty
}P(r)=0$.

In the limit of small $r$, corresponding to small values of the parameter $u$%
, we have $r\approx \left( C_{2}/2\right) \Gamma (p)\left(
l/2\right) ^{-p}u^{s-p}$, giving
\begin{equation}
u\approx \left[ \frac{2}{C_{2}\Gamma (p)}\left( \frac{l}{2}\right) ^{p}%
\right] ^{\frac{1}{s-p}}r^{\frac{1}{s-p}},r\rightarrow 0.
\end{equation}

Therefore in the limit of small radial distances the metric coefficients $%
e^{\nu }$ and $e^{\lambda }$ behave as
\begin{equation}
e^{\nu }\approx C^{2}\left[ \frac{2}{C_{2}\Gamma (p)}\left( \frac{l}{2}%
\right) ^{p}\right] ^{\frac{2(1-2s)}{s-p}}r^{\frac{2(1-s-p)}{s-p}%
},r\rightarrow 0,
\end{equation}
and
\begin{equation}
e^{\lambda }\approx 3l^{2}\left[ \frac{2}{C_{2}\Gamma (p)}\left( \frac{l}{2}%
\right) ^{p}\right] ^{\frac{1}{s-p}}\frac{r^{\frac{2}{s-p}}}{\left\{ 1+\frac{%
\Gamma (p+1)}{\Gamma (p)}+\left( \frac{l}{2}\right) ^{2}\frac{\Gamma (p-1)}{%
\Gamma (p)}\left[ \frac{2}{C_{2}\Gamma (p)}\left( \frac{l}{2}\right) ^{p}%
\right] ^{\frac{2}{s-p}}r^{\frac{2}{s-p}}\right\}
^{2}},r\rightarrow 0,
\end{equation}
respectively.

In the limit of large $r$ the dependence of the metric functions
on the radial distance can be given only in a parametric form. Since for $%
u\rightarrow \infty $ we have $r=\left( C_{1}/\sqrt{2\pi l}\right)
u^{s-1/2}\exp (lu)$, the asymptotic behavior of $\exp \left( \nu
\right) $ and $\exp \left( \lambda \right) $ is given, for large
$r$, in the parametric form, with $u$ taken as parameter
\begin{equation}
r\approx \frac{C_{1}}{\sqrt{2\pi l}}u^{s-1/2}e^{lu},e^{\nu }\approx \frac{%
C^{2}C_{1}^{2}}{2\pi l}u^{1-2s}e^{2lu},r\rightarrow \infty ,
\end{equation}
and
\begin{equation}
r\approx \frac{C_{1}}{\sqrt{2\pi l}}u^{s-1/2}e^{lu},e^{\lambda
}\approx \frac{3l^{2}u^2}{\left( s+lu\right) ^{2}},r\rightarrow
\infty .
\end{equation}

For $r\rightarrow 0$ the radial distance dependence of the dark
radiation $U$ is given by
\begin{equation}
3\beta U(r)\approx \frac{1}{3r^{2}}\left( 1+\frac{K_{U}}{r^{\frac{2}{s-p}}}%
\right) ,r\rightarrow 0,
\end{equation}
where
\begin{equation}
K_{U}=\frac{2n^{2}+\left[ s+\Gamma (p+1)/\Gamma (p)\right]
^{2}+2(2s-1)\left[
s+\Gamma (p+1)/\Gamma (p)\right] }{l^{2}\left[ \frac{2}{C_{2}\Gamma (p)}%
\left( \frac{l}{2}\right) ^{p}\right] ^{\frac{2}{s-p}}}.
\end{equation}

In the limit of large $r$, $r\rightarrow \infty $, the dark
radiation behaves as
\begin{equation}
3\beta U(r)\approx \frac{2}{3r^{2}},r\rightarrow \infty .
\end{equation}

In the limit of small $r$ we find for the dark pressure
\begin{equation}
3\beta P(r)\approx -\frac{2}{r^{2}}\left( 1+\frac{K_{P}}{r^{\frac{2}{s-p}}}%
\right) ,r\rightarrow 0,
\end{equation}
where
\begin{equation}
K_{P}=\frac{n^{2}-4\left[ s+\Gamma (p+1)/\Gamma (p)\right]
^{2}+2(2s-1)\left[
s+\Gamma (p+1)/\Gamma (p)\right] }{6l^{2}\left[ \frac{2}{C_{2}\Gamma (p)}%
\left( \frac{l}{2}\right) ^{p}\right] ^{\frac{2}{s-p}}}.
\end{equation}

For $r\rightarrow \infty $ the dark pressure has the asymptotic
limit
\begin{equation}
3\beta P(r)\approx -\frac{1}{3r^{2}},r\rightarrow \infty .
\end{equation}

Therefore at infinity the dark pressure and the dark radiation
obey the equation of state
\begin{equation}
U+2P=0.
\end{equation}

\section{Stable circular orbits in vacuum brane space-times with
general conformal symmetry}

As a physical application of the conformally symmetric brane
metric (\ref {line}) generated by a vector field $\xi $ with
arbitrary symmetry we consider now the problem of constructing
stable circular timelike geodesic orbits. The motion of a test
particle in the gravitational field can be described by the
Lagrangian \cite{Ma03}
\begin{equation}
2L=\left( \frac{ds}{d\tau }\right) ^{2}=-e^{\nu \left( r\right)
}\left(
\frac{dt}{d\tau }\right) ^{2}+e^{\lambda \left( r\right) }\left( \frac{dr}{%
d\tau }\right) ^{2}+r^{2}\left( \frac{d\Omega }{d\tau }\right)
^{2},
\end{equation}
where we denoted by $\tau $ the affine parameter along the
geodesics. In the timelike case $\tau $ corresponds to the proper
time. The equation giving the tangential velocity of the body has
been derived in \cite{Ma03} and \cite{La03}, and is given by
\begin{equation}
v_{tg}^{2}=\frac{r\nu ^{\prime }}{2}.
\end{equation}

Thus, the rotational velocity of the test body is determined by
the metric coefficient $\exp \left( \nu \right) $ only.

In the case of the motion of a test particle in a conformally
symmetric, static spherically symmetric space-time, with a general
non-static vector field generating the symmetry, the metric
coefficient $\exp \left( \nu \right) $ is given by Eq. (\ref{nu}).
Therefore as a function of the conformal factor $\psi $ the
angular velocity $v_{tg}$ is given by the simple expression
\begin{equation}
v_{tg}^{2}=1-\frac{\alpha (2s-1)}{\psi -\alpha }.  \label{tg}
\end{equation}

Eq. (\ref{tg}) gives a simple physical interpretation of the
conformal factor $\psi $ in terms of the tangential velocity,
$\psi =\alpha \left[ 1+(2s-1)\left( 1-v_{tg}^{2}\right)
^{-1}\right] $. Hence the conformal factor is proportional to the
tangential velocity of a test particle in the gravitational field
of a brane with general conformal symmetry. As a function of the
parameter $u$, $v_{tg}$ is given by
\begin{equation}
v_{tg}^{2}=1-\frac{2s-1}{u\frac{d}{du}\ln r}.  \label{v1}
\end{equation}

Together with Eq. (\ref{r}), Eq. (\ref{v1}) gives the parametric
representation of the tangential velocity of a test particle in a
stable circular orbit on the radial distance $r$. By using the
parametric dependence of $r$ we obtain the following explicit
representation for $v_{tg}$ as a function of $u$:
\begin{equation}
v_{tg}^{2}\left( u\right)
=1-\frac{2s-1}{s+\frac{lu}{2}\frac{C_{1}\left[ I_{p-1}\left(
lu\right) +I_{p+1}(lu)\right] +C_{2}\left[ K_{p-1}\left( lu\right)
+K_{p+1}(lu)\right] }{C_{1}I_{p}\left( lu\right)
+C_{2}K_{p}(lu)}}.
\end{equation}

The variation of $v_{tg}$ as a function of the radial distance $r$
is represented, for different values of the constant $s$, in Fig.
5.

\vspace{0.2in}
\begin{figure}[h]
\includegraphics{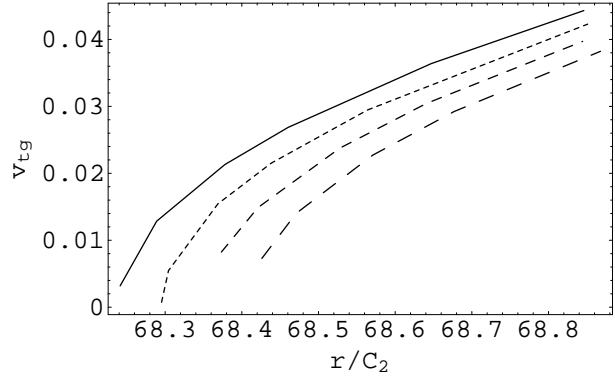}
\caption{Variation, as a function of the parameter $r/C_2$, of the
tangential velocity $v_{tg}$ of a test particle in a stable
circular orbit in a conformally symmetric vacuum space-time with a
general conformal vector on the brane, for $l=10^{-3.2}$,
$R_0=C_1/C_2=10^5$ and different values of $s$: $s=1.1$ (solid
curve), $s=1.101$ (dotted curve), $s=1.102$ (dashed curve) and
$s=1.103$ (long dashed curve). } \label{FIG5}
\end{figure}

The tangential velocity of a particle in a stable circular orbit
in a conformally symmetric geometry on the brane, with arbitrary
conformal factor, is a monotonically increasing function of the
distance.

In the limit of small radial distances, by using the approximate
expressions of the Bessel functions for small values of the
argument we obtain the following explicit dependence of the
tangential velocity on the radial distance $r$:
\begin{equation}
v_{tg}^{2}\left( r\right) \approx 1-\frac{2s-1}{s+\frac{\Gamma
\left(
p+1\right) }{\Gamma \left( p\right) }+\left( \frac{2}{C_{2}}\right) ^{\frac{2%
}{s-p}}\left( \frac{l}{2}\right) ^{\frac{s+p}{s-p}}\Gamma (p-1)\Gamma ^{%
\frac{p-s-2}{s-p}}\left( p\right) r^{\frac{2}{s-p}}},r\rightarrow
0.
\end{equation}

The angular velocity is a monotonically  increasing function of
$r$. For small values of $r$, one can neglect the term containing
the radial coordinate with respect to the other terms, giving
$v_{tg}^{2} \approx 1-(2s-1)/\left[s+\Gamma (p+1)/\Gamma
(p)\right]=$constant. Therefore at small distances from the origin
the angular velocity of a test particle moving on the conformally
symmetric brane has a constant value.

In the limit of large distances  we obtain the following
parametric representation for $v_{tg}$:
\begin{equation}
r\approx \frac{C_{1}}{\sqrt{2\pi l}}u^{s-1/2}e^{lu},v_{tg}^{2}(u)\approx 1-%
\frac{2s-1}{s+lu}.
\end{equation}

In the limit of large $r$ the tangential velocity of a test
particle in a stable circular orbit in a vacuum brane with general
conformal symmetry tends to the speed of light $c$. This type of
behavior, corresponding to an increase in the velocity of test
particles in a stable circular orbit,  is very similar to the
behavior of hydrogen clouds outside spiral galaxies, and which is
usually attributed to the presence of the dark matter. Therefore,
based on this analogy,  we shall assume that the conformally
symmetric solution of the gravitational field equations on the
brane describes the gravitational field outside a galaxy.

The metric coefficient $\exp (\lambda )$ can be expressed as a
function of the tangential velocity and the parameter $u$ as
\begin{equation}
e^{\lambda }=3\left( \frac{l}{2s-1}\right) ^{2}u^{2}\left(
1-v_{tg}^{2}\right) ^{2}.
\end{equation}

The field equation (\ref{f1}) can be immediately integrated to
give the following representation for the metric tensor component
$\exp \left( -\lambda \right) $,
\begin{equation}
e^{-\lambda }=1-\frac{2GM_{U}(r)}{r},
\end{equation}
where
\begin{equation}
M_{U}(r)=\frac{3\beta }{2G}\int_{0}^{r}U(r)r^{2}dr,
\end{equation}
is the mass corresponding to the dark radiation term $U(r)$ (the
''dark mass''). By using Eq. (\ref{lambda}) we obtain the
following parametric representation for $M_{U}(r)$:
\begin{equation}
M_{U}(r)=\frac{r}{2G}\left[ 1-\frac{1}{3l^{2}}\left(
\frac{d}{du}\ln r\right) ^{2}\right] .
\end{equation}

The dark mass can be also expressed in terms of the tangential
velocity of a particle in a stable circular orbit as
\begin{equation}
M_{U}(r)=\frac{r}{2G}\left[ 1-\frac{\left( 2s-1\right) ^{2}}{3l^{2}}\frac{1}{%
u^{2}\left( 1-v_{tg}^{2}\right) ^{2}}\right] .
\end{equation}

In the limit of small $r$ the dark mass behaves like
\begin{equation}
M_{U}(r)\approx \frac{r}{2G}\left\{ 1-\frac{1}{3l^{2}}\left[
s+\frac{\Gamma
(p+1)}{\Gamma (p)}\right] ^{2}\left[ \frac{2}{C_{2}\Gamma (p)}\left( \frac{l%
}{2}\right) ^{p}\right]
^{-\frac{2}{s-p}}\frac{1}{r^{\frac{2}{s-p}}}\right\} ,r\rightarrow
0.
\end{equation}

In the limit of large $r$ ($u\rightarrow \infty $) we obtain
\begin{equation}\label{mass}
M_{U}(r)=\frac{r}{3G}.
\end{equation}

Therefore for large $r$ the dark mass is linearly increasing with
the distance to the galactic center.

\section{Discussions and final remarks}

In the present paper we have obtained a class of conformally
symmetric solutions of the vacuum field equations in the brane
world model, under the assumption of a non-static and non-central
conformal symmetry, and we have discussed some of their physical
properties. To obtain the solution we have made the crucial
assumption of ignoring the possible effect of the cosmological
constant on the geometrical structure of the vacuum on the brane.
Mathematically, this means that the model can correctly describe
the dynamics of particles at distances $r$ from the galactic
center satisfying the condition
\begin{equation}
r<<\sqrt{\frac{l^{2}u^{2}-n^{2}}{12l\Lambda u^{2}}}.
\end{equation}

Taking into account the smallness of the value of the cosmological
constant, the present model can be safely applied to describe the
motion of particles outside galaxies. On the other hand, in the
limit of large $r$, both the dark radiation $U$ and the dark
pressure $P$ are decreasing functions of $r$, and at infinity the
geometry and the dynamical behavior of the particles is determined
by the cosmological constant only.

The behavior of the metric coefficients and of the angular
velocity in the solution we have obtained depends on four
arbitrary constants of integration $l$, $s$, $C_{1}$ and $C_{2}$.
Their numerical values can be obtained by assuming the continuity
of the metric coefficients $\exp \left( \lambda \right) $ and
$\exp \left( \nu \right) $ across the vacuum boundary of the
galaxy. For simplicity we assume that inside the ''normal''
(baryonic) luminous matter, with density $\rho _{B}$, which form a
galaxy, the non-local effects of the Weyl tensor can be neglected.
We define the vacuum boundary $r_{0}$ of the galaxy (which, for
simplicity, is assumed to have
spherical symmetry) by the condition $\rho _{B}\left( r_{0}\right) \approx 0$%
.

Therefore at the vacuum boundary the metric coefficients are given
by $\exp\left( \nu \right) =\exp \left( -\lambda \right) =1-2GM_{B}/r_{0}$, where $%
M_{B}=4\pi \int_{0}^{r_{0}}\rho _{B}\left( r\right) r^{2}dr$ is
the total baryonic mass inside the radius $r_{0}$. For simplicity
we shall also assume that the tangential velocity at the galactic
boundary can be approximated by its Newtonian expression,
$v_{tg}^{2}\left( r_{0}\right) \approx GM_{B}/r_{0} $.

The continuity of the radius $r$ through the surface $r=r_{0}$
gives the first condition which must be satisfied by the
integration constants:
\begin{equation}
r_{0}=u_{0}^{s}\left[ C_{1}I_{p}\left( lu_{0}\right)
+C_{2}K_{p}\left( lu_{0}\right) \right] ,
\end{equation}
where $u_{0}$ is the value of the parameter $u$ corresponding to
the vacuum boundary of the galaxy $r=r_{0}$. The continuity of
$\exp \left( \nu \right) $ at $r=r_{0}$ gives
\begin{equation}
C^{2}r_{0}^{2}u_{0}^{2\left( 1-2s\right) }=1-2GM_{B}/r_{0}.
\end{equation}

The continuity of the tangential velocity of a test particle
across the galactic boundary fixes the value of the derivative
$dr/du$ for $u=u_{0}$:
\begin{equation}
\left( \frac{dr}{du}\right) _{u=u_{0}}=\frac{r_{0}}{u_{0}}\frac{2s-1}{1-%
\frac{GM_{B}}{r_{0}}}.
\end{equation}

Then the continuity of $\exp \left( \lambda \right) $ gives
\begin{equation}
u_{0}^{2}\left( 1-\frac{GM_{B}}{r_{0}}\right) ^{2}\left(
1-2GM_{B}/r_{0}\right) =\frac{1}{3}\left( \frac{2s-1}{l}\right)
^{2}.
\end{equation}

Thus the continuity of the metric potentials gives the following
compatibility condition, which must be satisfied by the constants
$C$, $l$ and $s$:
\begin{equation}
C^{2}\left[ \frac{1}{3}\frac{2s-1}{l}\right] ^{2(1-2s)}=\frac{1}{r_{0}^{2}}%
\left( 1-\frac{GM_{B}}{r_{0}}\right) ^{2(1-2s)}\left( 1-\frac{2GM_{B}}{r_{0}}%
\right) ^{1-2s}.
\end{equation}

As a possible physical application of the obtained solutions we
have considered the behavior of the angular velocity of a test
particle in a stable circular orbit on the brane with non-static
and non-central symmetry. The conformal factor $\psi $, together
with two constants of integration, uniquely determines the
rotational velocity of the particle. The angular velocity is
always an increasing function of the radial distance $r$, which
from a physical point of view can be considered as the distance
from the galactic center. This behavior is independent on the
numerical values of the parameters (separation and integration
constants) of the model. In the limit of large radial distances
the angular velocity tends to a constant value, which in the
present case is the speed of light $c$. This general behavior is
typical for massive particles (hydrogen clouds) outside galaxies.
Thus the rotational galactic curves can be naturally explained in
brane world models, without using the concept of dark matter.

It has long been known that Newtonian or general relativistic
mechanics applied to the visible matter in galaxies and clusters
does not correctly describe the dynamics of those systems. The
rotation curves of spiral galaxies \cite{Bi87} are one of the best
evidences showing the problems Newtonian mechanics and/or standard
general relativity has to face on the galactic/intergalactic
scale. In these galaxies, neutral hydrogen clouds are observed at
large distances from the center, much beyond the extent of the
luminous matter. Assuming a non-relativistic Doppler effect and
emission from stable circular orbits in a Newtonian gravitational
field, the frequency shifts in the $21$ cm line hydrogen emission
lines allows the measurement of the velocity of the clouds.

Observations show that the rotational velocities increase near the
center of the galaxy and then remain nearly constant at a value of
$v_{tg\infty }\sim 200$ km/s \cite{Bi87}. This leads to a mass
profile $M(r)=rv_{tg\infty }^2/G$. Consequently, the mass within a
distance $r$ from the center of the galaxy increases linearly with
$r$, even at large distances where very little luminous matter can
be detected. This behavior of the galactic rotation curves is
explained by postulating the existence of some dark (invisible)
matter, distributed in a spherical halo around the galaxies. The
dark matter is assumed to be a cold, pressureless medium. There
are many possible candidates for dark matter, the most popular
ones being the weekly interacting massive particles (WIMP).

However, despite more than 20 years of intense experimental and
observational effort, up to now no {\it non-gravitational}
evidence for dark matter has ever been found: no direct evidence
of it and no annihilation radiation from it. Moreover, accelerator
and reactor experiments do not support the physics (beyond the
standard model) on which the dark matter hypothesis is based.

Therefore, it seems that the possibility that Einstein's (and the
Newtonian) gravity breaks down at the scale of galaxies cannot be
excluded {\it a priori}. Several theoretical models, based on a
modification of Newton's law or of general relativity, have been
proposed to explain the behavior of the galactic rotation curves
\cite{dark}. In the framework of the brane models, the role of the
dark matter is played by the dark mass $M_U(r)$, which is the mass
associated to the dark radiation term, having its physical origin
in the five-dimensional bulk. As can be seen from Eq.
(\ref{mass}), the dark mass is indeed linearly increasing with the
distance to the galactic center. Thus the dark mass has a similar
behavior to that observed in the case of the "dark matter" around
the galaxies.

The main advantage of the brane world models as compared to the
alternative explanations of the galactic rotation curves and of
the dark matter is that it can give a systematic and coherent
description of the Universe from galactic to cosmological scales.
On the other hand, in the present model all the relevant physical
quantities, including the dark energy and the dark pressure, which
describe the non-local effects due to the gravitational field of
the bulk, are expressed in terms of observable parameters (the
baryonic mass and the radius of the galaxy). Therefore this opens
the possibility of the testing of the brane world models by using
astronomical and astrophysical observations at the galactic scale.
A systematic comparison between the predictions of the model
discussed in the present paper and the observational results will
be considered in detail in a future publication.

\section*{Acknowledgements}

This work was supported by a Seed Funding Programme for Basic
Research of the Hong Kong Government.

\end{document}